\documentclass[onecolumn, draftcls]{IEEEtran}

\usepackage{graphicx}
\usepackage{color}
\usepackage{amsmath}
\usepackage{amsfonts}
\usepackage{epsfig}
\usepackage{epstopdf}
\usepackage{amssymb}
\graphicspath{{./Figures/}}

\usepackage{paralist}

\usepackage{subcaption}
\usepackage[list=true]{subcaption}
\usepackage{cite}

\newfont{\bbb}{msbm10 scaled 500}

\newfont{\bb}{msbm10 scaled 1100}

\newcommand{\gv}{{\bf g}}
\newcommand{\hv}{{\bf h}}

\newcommand{\vv}{{\bf v}}
\newcommand{\xv}{{\bf x}}

\newcommand{\Am}{{\bf A}}

\newcommand{\Hm}{{\bf H}}

\newcommand{\Km}{{\bf K}}

\newcommand{\Um}{{\bf U}}

\newcommand{\Vm}{{\bf V}}

\newcommand{\Ac}{{\cal A}}
\newcommand{\Bc}{{\cal B}}
\newcommand{\Cc}{{\cal C}}

\newcommand{\Ec}{{\cal E}}

\newcommand{\Kc}{{\cal K}}
\newcommand{\Lc}{{\cal L}}
\newcommand{\Mc}{{\cal M}}
\newcommand{\Nc}{{\cal N}}

\newcommand{\Rc}{{\cal R}}

\newcommand{\Sigmam}{\hbox{\boldmath$\Sigma$}}

\definecolor{OXO-emph}{RGB}{153,0,0}

\allowdisplaybreaks

\begin{document}
%
\title{Exploiting Full-duplex Receivers for Achieving Secret Communications in Multiuser MISO Networks}

\author{\IEEEauthorblockN{Berk Akgun, O. Ozan Koyluoglu, and Marwan Krunz
\IEEEcompsocitemizethanks{\IEEEcompsocthanksitem The authors are with the Department
of Electrical and Computer Engineering, University of Arizona, Tucson,
AZ, 85721.\protect\\
E-mail: \{berkakgun, ozan, krunz\}@email.arizona.edu }
\IEEEcompsocitemizethanks{\IEEEcompsocthanksitem Digital Object Identifier 10.1109/TCOMM.2016.2641949 }
\IEEEcompsocitemizethanks{0090-6778~\copyright~2016 IEEE. Personal use is permitted, but republication/redistribution requires IEEE permission.}
}
}

\maketitle

\begin{abstract}
We consider a broadcast channel, in which a multi-antenna transmitter (Alice) sends $K$ confidential information signals to $K$ legitimate users (Bobs) in the presence of $L$ eavesdroppers (Eves).
Alice uses MIMO precoding to generate the information signals along with her own (Tx-based) friendly jamming. Interference at each Bob is removed by MIMO zero-forcing. This, however, leaves a ``vulnerability region'' around each Bob, which can be exploited by a nearby Eve.
We address this problem by augmenting Tx-based friendly jamming (TxFJ) with Rx-based friendly jamming (RxFJ), generated by each Bob.
Specifically, each Bob uses self-interference suppression (SIS) to transmit a friendly jamming signal while simultaneously receiving an information signal over the same channel. We minimize the powers allocated to the information, TxFJ, and RxFJ signals under given guarantees on the individual secrecy rate for each Bob.
The problem is solved for the cases when the eavesdropper's channel state information is known/unknown. Simulations show the effectiveness of the proposed solution. Furthermore, we discuss how to schedule transmissions when the rate requirements need to be satisfied on average rather than instantaneously. Under special cases, a scheduling algorithm that serves only the strongest receivers is shown to outperform the one that schedules all receivers.
\end{abstract}


\section{Introduction}

As wireless systems continue to proliferate, confidentiality of their communications becomes one of the main concerns due to the broadcast nature of the wireless medium. Cryptographic techniques can be utilized to address these concerns, but such techniques often assume adversaries with limited computational capabilities. Physical (PHY) layer security, on the other hand, can be implemented regardless of the adversary's computational power. It also takes advantage of the characteristics of the wireless medium.

\subsection{Related Work}

The origins of PHY-layer security dates back to the pioneering work of Wyner \cite{Wyner:wire75} that studied the concept of \textit{secrecy capacity} for the degraded wiretap channel.
The authors in \cite{Csiszar:Broadcast78} extended Wyner's work to non-degraded discrete memoryless broadcast channels.
Later on, the secrecy capacity of MIMO (multiple-input multiple-output) wiretap channel was obtained in \cite{Oggier:secrecy11}.
The secrecy region of the Gaussian MIMO broadcast channel was studied in \cite{Bagherikaram:secrecy13}, \cite{Ekrem:Capacity12}, and \cite{Liu:MIMO10}. 
The authors in \cite{Chen2015} and \cite{Chen2015a} studied the problem of secure communications over broadcast channels under individual secrecy constraint, which guarantees that the information leakage to eavesdroppers from each information message vanishes. Even though the joint secrecy constraint, which ensures that the information leakage to eavesdroppers from all information messages vanishes, is stronger than the individual one, it is not always possible to satisfy.
Moreover, the individual secrecy constraint still offers an acceptable secrecy level, while increasing transmission rates \cite{Chen2015}.
To facilitate secrecy, Goel and Negi \cite{Goel:Guaranteeing08} introduced the concept of artificial noise, a.k.a. \textit{friendly jamming} (FJ), for Gaussian channels.
The idea is to artificially generate Gaussian noise over the channel in order to degrade eavesdropping.
The legitimate receivers remain unaffected, as the FJ signals are generated to be orthogonal to their channels, utilizing the MIMO precoder techniques.
This is a special case of the channel prefixing technique proposed in \cite{Csiszar:Broadcast78}, which randomizes the codewords before sending them over the channel.
The authors in \cite{Yang:Joint14} studied a multiuser broadcast channel where a sender transmits $K$ independent streams to $K$ receivers.
Linear precoding and FJ techniques were proposed to enhance PHY-layer security.
The authors in \cite{Romero-Zurita:Outage12} studied an outage probability based power allocation problem for data and FJ so as to satisfy certain secrecy requirements.
A full-duplex (FD) receiver that sends FJ to secure the communication was considered in \cite{Li:Secure12} and \cite{Romero-Zurita:Can15}.
Their work was later extended to allow both transmitter and receiver to generate FJ in \cite{Li:Secure14}.
In that model, at least two antennas were needed at the receiver, one for sending the FJ signal and the other to receive the information message.
A similar system model was used in \cite{Zhou:Securing13} where one of the antennas at the receiver is utilized to receive information signals, and the remaining ones generate FJ signals.
The authors in \cite{Zhou2014} extended this work to a MIMO system where $r$ antennas of the receiver are selected to receive information signals, while the remaining antennas generate FJ signals. 
The authors in \cite{Zheng:Improving13} showed that PHY-layer secrecy can be enhanced using FD jamming receivers without assuming perfect self-interference suppression (SIS).
Another system model with one FD base station (BS), one transmitter, one receiver, and one eavesdropper was considered in \cite{Zhu:Joint14}.
In this model, the BS receives a message from the transmitter, while sending an information message to the receiver together with an FJ signal.
It was assumed that the transmitter's signal does not interfere at the receiver, and the problem of maximizing the secret transmission rate was investigated.
This work was extended to a multiuser communication system with multiple single-antenna uplink and downlink users and multi-antenna eavesdoppers by the authors in \cite{Sun2015}.
They formulated a multi-objective optimization problem to minimize the total downlink and uplink transmit power, while guarenteeing both uplink and downlink security. In their model, the only FJ source was the FD BS, and zero-forcing beamforming was employed for  uplink transmissions. 
The authors in \cite{Feng2016} considered an FD two-way secure communication system where two FD sources are equipped with multiple transmit antennas and a single receive antenna in the presence of a single-antenna eavesdropper. Remarkably, the sources in this model do not employ FJ signals to further impair the eavesdropper's channel.
None of these works considered a multiuser scenario where multiple receivers generate FJ signals.
In contrast, here, we consider a $K$-user scenario with single-antenna FD receivers that generate FJ signals.
Furthermore, we consider a scenario where the information signals should not be decoded at unintended receivers (confidential messages), and the information leakage to eavesdroppers (they may be system devices that are not necessarily malicious, but ``untrusted") for each information message should vanish (individual secrecy).
Multiuser broadcast channels even without any FJ signal lead to non-convex problem formulations due to interference from unintended information signals.
When FJ signals are incorporated to the system to provide secure communications against eavesdroppers, the problem becomes harder to deal with.
The joint power allocation among FJ signals (that are generated both at the transmitter and $K$ receivers) and information signals for $K$ simultaneous transmissions has not been explored previously. In addition, we specifically focus on a problem which arises from eavesdroppers whose channels are correlated with those of legitimate receivers.
A similar system model was considered in \cite{akgun2015receiver}, where the BS sends two independent data streams to only two legitimate receivers in the presence of a multi-antenna eavesdropper.
In that work, we developed a secrecy encoding scheme to construct the information signals under joint secrecy constraints.
We also characterized the achievable sum-rate, and investigated a special case of the corresponding optimization problem.

\subsection{Motivation and Contributions}

Our work is motivated by recent studies that showed the vulnerability of the intended receiver to adversaries in its proximity \cite{He:security} and \cite{He:Is}.
In particular, when the eavesdropper's channel is highly correlated with that of a legitimate receiver, MIMO-based nullification of the transmitter's FJ signal at that receiver, a.k.a. zero-forcing beamforming (ZFBF), extends to nearby eavesdroppers.
This increases the signal-to-interference-plus-noise ratio (SINR) at the eavesdroppers (Eves), significantly reducing the secrecy rate.
The goal of our work is to provide message confidentiality, independent of the amount of correlations between the channel state information (CSI) at Eve and the intended receiver.
We consider a scenario where the transmitter (Alice) sends $K$ independent confidential messages to $K$ legitimate receivers (Bobs).
To achieve message confidentiality, we propose to use receiver-based friendly jamming (RxFJ) along with transmitter-based friendly jamming (TxFJ).
This way, Eve's received signal is degraded even if her CSI is highly correlated with those of Bobs. To remove TxFJ at each Bob, ZFBF is employed by Alice.
This technique also provides confidentiality for the information messages (information signals are zero-forced at unintended receivers). Even though ZFBF technique is a suboptimal solution for broadcast channels, it significantly reduces the implementation complexity \cite{Spencer:Zero04, Yoo:optimality06}.
In fact for multiuser MIMO channels, ZFBF is asymptotically optimal in high SNR regimes, e.g. $10$ dB, and in some low SNR regimes, in terms of throughput maximization as well as power minimization \cite{Spencer:Zero04}. Moreover, as the number of users becomes very large, the sum-rate performance of ZFBF is close to optimal \cite{Yoo:optimality06}.

We formulate an optimization problem to minimize the total power consumption for the information, TxFJ, and RxFJ signals, while guaranteeing a certain individual secrecy rate for each Bob with/without Eve's CSI (ECSI).
(In unknown ECSI case, it is assumed that the first- and the second-order statistics of ECSI are known).
We exploit the conditions where using RxFJ together with TxFJ has better system performance than using only TxFJ or ZFBF in terms of preventing information leakage to Eves. 

The contributions of this paper can be summarized as follows:
\begin{itemize}
\item We show that FD capabilities can be exploited in multiuser MISO (multiple-input single-output) networks to provide confidential communications using RxFJ against eavesdroppers, whose channels are correlated with that of legitimate receivers.  
\item We investigate the joint power allocation problem for information, TxFJ, and RxFJ signals to satisfy certain secrecy rate requirements, and provide optimal solutions for practical systems.
\item We determine the optimal randomization rates for wiretap coding to confuse the eavesdroppers based on the given requirements (individual secrecy rate requirement if ECSI is known, and secrecy outage probability requirement if only the statistics of ECSI are known).
\item We analyze the effect of different scheduling approaches on the performance of the proposed schemes. 
\end{itemize} 

The rest of the paper is organized as follows. Section II describes the system model. In Section III, we present different beamforming (BF) techniques for scenarios with known/unknown ECSI. Optimization problem is formulated and analyzed in Section IV. We provide simulation results and propose a scheduling scheme in Section V. The paper is concluded in Section VI.
(The published version of this work can be found in \cite{berk2016journal}.)

Throughout the paper, we adopt the following notation. Vectors and matrices are denoted by bold lower-case and upper-case letters, respectively. We use column and row vectors notations interchangeably. $ (\cdot)^{*} $ and $ (\cdot)^{T} $ represent the complex conjugate transpose and the transpose of a vector or matrix, respectively. Frobenius norm and the absolute value of a real or complex number are denoted by $ \Vert \cdot \Vert $ and $ \vert \cdot \vert $, respectively. $ \mathbb{E}[\cdot] $ indicates the expectation of a random variable. $ \Am \in \mathbb{C}^{M \times N} $ means that $ \Am $ is an $ M \times N $ complex matrix.
$ \Cc\Nc(\mu, \sigma^{2}) $ denotes complex circularly symmetric Gaussian random variable with mean $ \mu $ and variance $ \sigma^{2} $.
$ \textbf{I}_{N} $ represents an $ N \times N $ identity matrix.
$[x]^{+}= \mathrm{max}(x, 0)$.
$\mathrm{rank(\Am)}$ indicates the rank of matrix $\Am$.
$I ( X; Y)$ refers to the mutual information between random variables $X$ and $Y$.
Let $\Ac$ and $\Bc$ be two sets.
Then, $\{ \Ac \setminus \Bc \}$ indicates the set of all elements of $\Ac$ that are not in $\Bc$.
For simplicity, $\log_2(.)$ is referred to as $\log(.)$ in the rest of the paper.

\section{System Model}
\label{sec:system}

\begin{figure}[!t]
\centering
\includegraphics[width=95mm]{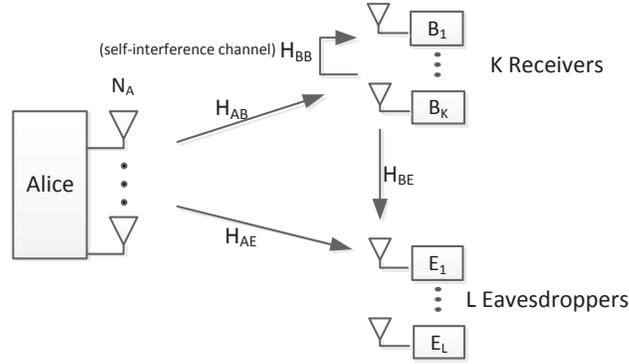}
\caption{MU-MISO system model with both TxFJ and RxFJ.}
\label{fig:scenario1}
\end{figure}

As shown in Figure \ref{fig:scenario1}, we consider a MU-MISO (multiuser MISO) network in which Alice transmits $K$ independent confidential data streams to $K$ receivers in the presence of $L$ eavesdroppers.
$\Bc = \{ B_1, B_2, \cdots, B_K\}$ is the set of legitimate receivers, each has a single-antenna FD radio (the same antenna is used to simultaneously transmit and receive signals over the same frequency) \cite{Bharadia:Full13}.
$\Ec = \{ E_1, E_2, \cdots, E_L\}$ is the set of eavesdroppers, each of which has a single-antenna.
Legitimate receivers and eavesdroppers are referred to as Bobs and Eves in the rest of the paper, respectively.
Let the number of antennas at Alice be $ N_A $.
Let $ \xv_{A} \in  \mathbb{C}^{N_{A} \times 1} $ be Alice's transmit signal.
$ x_{B_k} $ denotes the transmit signal from the $k$th Bob, $B_k$, $\forall k \in \Kc$, where $\Kc$ represents the set $\{1, \cdots, K\}$ throughout the paper, i.e., $\Kc = \{1, \cdots,K\}$.
Similarly, $\Lc$ denotes the set $\{1, \cdots, L\}$ in the rest of the paper.

The signals received by the $k$th Bob and the $i$th Eve at time $ t \in \{ 1, \cdots, n \}$ are, respectively, given by:	
    \begin{align}   
	y_{B_k}^t &= \hv_{AB_k} \xv_{A}^t + \sqrt{\alpha} h_{B_kB_k}x_{B_k}^t + \sum_{l \in \{\Kc \setminus k\}} h_{B_lB_k}x_{B_l}^t + n_{B_k}^t, \;\forall k \in \Kc\\
	z_{E_i}^t &= \hv_{AE_i} \xv_{A}^t + \sum_{k \in \Kc} h_{B_kE_i} x_{B_k}^t + n_{E_i}^t, \;\forall i \in \Lc
  	\end{align} 	
where $ \hv_{AB_k} \in \mathbb{C}^{1 \times N_{A}} $ is the channel vector between Alice and the $k$th Bob $\forall k \in \Kc$, while $ \hv_{AE_i} \in \mathbb{C}^{1 \times N_{A}} $ is the channel vector between Alice and the $i$th Eve $\forall i \in \Lc$.
$ h_{B_kE_i} $ denotes the channel between the $k$th Bob $\forall k \in \Kc$ and the $i$th Eve $\forall i \in \Lc$. $ h_{B_kB_k} $ and $ h_{B_lB_k} $ represent the self-interference channel at the $k$th Bob $\forall k \in \Kc$ and the channel between the $l$th and $k$th Bob $\forall l \in \{ \Kc \setminus k\}$, respectively.
The channel $\hv_{ij}$, $i \in \{ A \cup \Bc \}$ and $j \in \{ \Bc \cup \Ec \}$, is equal to $\sqrt{D_{ij}} \gv_{ij}$, where $D_{ij}$ and $\gv_{ij} \sim \Cc \Nc (0, \textbf{I}_{N_A})$ represent the path loss component and small-scale effects of the channel, respectively.
Since FD radio design is considered at the receivers, a residual self-interference term is incorporated into the model.
This residual term defines the portion of the self-interference left after suppression, and is denoted with the scale factor $ \alpha \in [0,1]$, e.g. $\alpha = 0$ means full-suppression (ideal case).
$ n_{B_k} \sim \Cc\Nc(0, N_0) $ and $ n_{E_i} \sim  \Cc\Nc(0, N_0) $ represent AWGN (Additive White Gaussian Noise) at the $kth$ Bob and the $i$th Eve, respectively. 

We impose the following instantaneous power constraints:
    \begin{align}
	\mathbb{E}[\xv_{A}^{*}\xv_{A}] &\leq \bar{P}_A   \label{eq:powerconsttt}  \\
	\mathbb{E}[\, \vert x_{B_k} \vert^{2} ] &\leq \bar{P}_{B_k}, \; \forall k \in \Kc
	\label{eq:powerconst}
  	\end{align} 
where $\bar{P}_{A}$ and $\bar{P}_{B_k}$'s are given constants. 	

An achievable individual secrecy rate tuple is defined as $\Rc = (R_1,R_2, \cdots, R_K)$ if there exists codebooks $(2^{nR_k}, n)$ which satisfy both the reliability and security constraints. Let $W_k$ define the secure message from Alice to the $k$th Bob $B_k$ where $W_k \in \mathcal{W}_k = [1 : 2^{nR_k}]$. The reliability of the transmission is given as:
\begin{eqnarray}
\Pr( \hat{W_k} \neq W_k) \leq \epsilon_0
\label{eq:relcons}
\end{eqnarray}
where $\epsilon_0 \rightarrow 0$ as $n \rightarrow \infty$, and  $\hat{W_k}$ is the estimated message at $B_k$. Let $\mathrm{Y}^{n}_{B_l}$ and $ \mathrm{Z}^{n}_{E_i}$ define the received signal sequences in $n$ channel uses at the $l$th Bob and the $i$th Eve, respectively. Accordingly, the individual secrecy constraints at Bobs and Eves are given by:
\begin{align}
I(W_{k}; \mathrm{Y}^{n}_{B_l}) &\leq \epsilon_1, \; \forall (k,l) \in (\Kc \times \{ \Kc \setminus k \})
\label{eq:secons1}\\
I(W_{k}; \mathrm{Z}^{n}_{E_i} ) &\leq \epsilon_2, \; \forall (k,i) \in ( \Kc \times \Lc)
\label{eq:secons2}
\end{align}
where $\epsilon_1 \rightarrow 0$ and $\epsilon_2 \rightarrow 0$ as $n \rightarrow \infty$.
The first secrecy constraint, (\ref{eq:secons1}), ensures the message confidentiality in an individual manner (a.k.a., individual secrecy), i.e., each information signal should have minimal leakage at unintended Bobs.
The second one, (\ref{eq:secons2}), provides the individual secrecy against external Eves.
Note that the individual secrecy constraints are considered throughout this paper rather than the joint secrecy constraints.
Furthermore, we consider a scenario where Eves do not collude.
Let $s_k^n$ represents the codeword in the codebook to be transmitted in $n$ channel uses. This signal has to contain enough randomness such that the mutual information leakage to Eves will vanish to satisfy (\ref{eq:secons2}). Therefore, the secret codebook is generated as follows. $2^{n(R_k + R_k^x)}$ sequences are independently generated according to a certain probability distribution, where $R_k^x$ defines the randomization rate.
Then, these sequences are distributed into $2^{nR_k}$ bins, where the bin index is defined by $W_k$.
Correspondingly, each bin has $2^{nR_k^x}$ codewords. Let $W_k^x$ define the index of the codewords in each bin.
As a result, each codeword is represented by two indices, i.e., $s_k^n(W_k,W_k^x)$.
In the rest of the paper, we will require $I(S_k;Y_{B_k}) \geq R_k + R_k^x$ to reliably decode secure message and randomization at $B_k$ $\forall k \in \Kc$, and $I(S_k; Z_{E_i}) \leq R_k^x$ $\forall (k, i) \in ( \Kc \times \Lc ) $ to achieve message security in the sense of individual secrecy.
(Note that the randomization decoding is necessary to remove ambiguity in the codewords to reveal the secret messages at Bobs. In addition, this adequate amount of randomization implies the security of the message. This is the well-known Wyner's wiretap code \cite{Wyner:wire75}, specialized to the individual secrecy notion studied in this paper.) The secrecy constraint (\ref{eq:secons1}), on the other hand, will be satisfied via ZFBF technique employed at Alice.

The general signaling scheme that we consider in this paper is given by:
  	\begin{align}
	\xv_{A}^t &= \sum_{k \in \Kc} \vv_{k}s_{k}^t(W_k, W_k^x) +  \sum_{m \in \Mc} \vv_{m}^{(j)} j_{m}^t, \quad t = \{1,2, \cdots, n\} 
	\label{eq:generalsignaling}
  	\end{align}
where $\Mc = \{ 1, \cdots, M \}$.
$ s_{k}^t \sim \Cc\Nc(0, P_{S_k}) $ is the information signal for the $k$th Bob at time $t$, and $ \vv_{k} \in \mathbb{C}^{N_{A} \times 1} $ is its normalized BF vector such that $ \vv_{k}^{*}\vv_{k} = 1 $. $ j_m^t \sim \Cc\Nc(0, P_{m}^{(j)}) $ and $ \vv_{m}^{(j)} \in \mathbb{C}^{N_{A} \times 1} $ are the $m$th TxFJ signal at time $t$ and its BF vector, respectively.
$M$ is the number of independent TxFJ signals, and it will be explained later in detail.
$ \vv_{m}^{(j)} $ is a unit vector as well.
The RxFJ signal transmitted by the $k$th Bob is given by $ x_{B_k} = j_{B_k} $, where $ j_{B_k} \sim \Cc\Nc(0, P_{B_k}), \; \forall k \in \Kc$.

\section{Beamforming Schemes}

In this section, we will discuss beamforming schemes that satisfy the individual secrecy constraints in (\ref{eq:secons1}) and (\ref{eq:secons2}).
We employ a well-known ZFBF technique, which allows to cancel out any signal at any receiver given its CSI, to prevent Bobs to decode the unintended information signals given that Alice knows CSI from all Bobs to herself.
For multiuser MIMO channels, ZFBF technique is asymptotically optimal both in the high SNR regimes, e.g. 10dB, and in some low SNR regimes in terms of throughput maximization as well as power minimization \cite{Spencer:Zero04}.
In addition, sum-rate performance of ZFBF is close to optimal, as the number of users is very large \cite{Yoo:optimality06}.
The performance gap between ZFBF and the optimal solution is, indeed, reduced when the confidentiality constraint is imposed (the capacity of a system with confidentiality constraints is less than the one without confidentiality constraints).
At the same time, employing ZFBF for multiuser MIMO channels reduces design complexity (even though the optimal precoding design is tractable for such systems, the matrix computations and iterations in the optimization process still cause a practicability concern for real time systems).
The inter-user interference is also removed by ZFBF.
That is, the multiuser MISO channel reduces to a single user MISO channel from the standpoint of each Bob (with the cost of reduced degree of freedom (DoF) for the information signals).
The authors in \cite{Shafiee2007} considered a MISO channel with a single antenna eavesdropper, and characterized the optimal achievable secrecy rates under the assumption that the Gaussian signaling is used for the information signals.
They concluded that \textit{``the beamforming direction of the information signals should be adjusted to be as orthogonal to the eavesdropping channel direction as possible, while being as close to the main channel direction as possible''}.
By inspiring from this result, we will investigate the following, possibly suboptimal, beamforming schemes to provide security against Eves.
In Section \ref{sec:ZF}, the information signals are transmitted to an orthogonal space to the CSI of the unintended Bobs and Eves, and no TxFJ/RxFJ signal is utilized.
On the other hand, in Section \ref{sec:FJwC}, we consider a beamforming scheme where the $k$th information signal $s_k^n(W_k, W_k^x)$ is transmitted in the direction of the $k$th Bob's channel, and TxFJ signals are utilized with and without the knowledge of ECSI.
Later in the sequel, we will discuss the vulnerabilities of Tx-based beamforming schemes when the channels of Bobs and Eves are correlated.
To do that, we utilize asymptotically optimal beamforming schemes with low complexity.
(Finding optimal beamforming schemes for the given setup is not the scope of this paper.)

\subsection{ZFBF}
\label{sec:ZF}

Here, it is assumed that the ECSI is known to Alice and Bobs.
This assumption will be discussed later in Section \ref{sec:SQoS}.
Without any friendly jamming signal, it is ensured that Eves are not able to receive any information regarding messages employing ZFBF.
All of the information signals are canceled out at Eves and unintended Bobs.
Therefore, security constraints given in (\ref{eq:secons1}) and (\ref{eq:secons2}) are satisfied, where $R_k^x$ is set to $0$ (no need to use randomization rate as Eves do not receive any information signals).
 Correspondingly, the transmit signal at Alice is given by:
\begin{align}
\xv_A^t = \sum_{k \in \Kc} \vv_k s_k^t(W_k, W_k^x)
\end{align}
To implement ZFBF technique, precoding vector, $\vv_k$, is designed such that Eves and Bobs except $B_k$ do not receive the information signal, $s_k$. Let us define the joint channel matrix from Alice to these receivers as
\begin{align}
\hat{\Hm}_{B_k} = [ \hv_{AB_1}^T \cdots \hv_{AB_{k-1}}^T \hv_{AB_{k+1}}^T \cdots \hv_{AB_K}^T \hv_{AE_1}^T \cdots \hv_{AE_L}^T ]^T.
\end{align}  
Let the singular value decomposition (SVD) of this matrix be $\hat{\Hm}_{B_k} = \hat{\Um}_{B_k} \hat{\Sigmam}_{B_k} \hat{\Vm}_{B_k}^* $.
(The SVD of an $m \times n$ matrix $\Am$ has a form $ \Um \Sigmam \Vm^* $, where $\Sigmam$ is an $ m \times n$ rectangular diagonal matrix with the singular values of $\Am$ on the diagonal.
$\Um$ and $\Vm$ are $m \times m$ and $n \times n$ unitary matrices, and the columns of these matrices are called the left- and right-singular vectors of $\Am$, respectively.)
We assume that $N_A > (L +K -1)$ (the number of columns of this matrix is larger than the number of its rows).
Let $U_1$ denote $\mathrm{rank}(\hat{\Hm}_{B_k})$. ($U_1 = (L + K-1)$ if $\hat{\Hm}_{B_k}$ is a full-rank matrix.)
Let $\hat{\Vm}_{B_k}^{(2)}$  correspond to the last $(N_A - U_1)$ columns of $\hat{\Vm}_{B_k}$.
Then, $\hat{\Vm}_{B_k}^{(2)}$ forms an orthogonal basis for the null space of $\hat{\Hm}_{B_k}$.
Using this decomposition, we set the precoder as $\vv_k = \hat{\Vm}_{B_k}^{(2)}\vv_k^{(2)}$ ($\vv_k^{(2)}$ will be explained shortly).
This way, Eves and the unintended Bobs will not be able to receive $s_k$, since it will be nullified at them, i.e., $\hat{\Hm}_{B_k} \vv_k = \mathbf{0}$ where $\mathbf{0}$ is a zero vector.
On the other hand, the new channel seen by the receiver $B_k$ becomes $(\hv_{AB_k} \hat{\Vm}_{B_k}^{(2)}) \in \mathbb{C}^{1 \times (N_A - U_1)}$. Note that this reduces to an interference free channel. To maximize the received signal power over this channel, the second part of the precoder (i.e., $\vv_k^{(2)}$) should be designed as follows.
Let the SVD of the new channel vector be $(\hv_{AB_k} \hat{\Vm}_{B_k}^{(2)}) = \Um_{new} \Sigmam_{new} \Vm_{new}^*$.
The first column of $\Vm_{new}$ forms an orthogonal basis for the range space of the new channel. Consequently, the second part of the precoder is chosen in this range space. Indeed, this vector is given by the following equation:  
\begin{align}
\vv_k^{(2)} = \frac{(\hv_{AB_k}\hat{\Vm}_{B_k}^{(2)})^*}{ \Vert \hv_{AB_k}\hat{\Vm}_{B_k}^{(2)} \Vert}
\end{align}
Overall, the precoding vector is designed as:
\begin{eqnarray}
\vv_k = \hat{\Vm}_{B_k}^{(2)} \frac{(\hv_{AB_k}\hat{\Vm}_{B_k}^{(2)})^*}{ \Vert \hv_{AB_k}\hat{\Vm}_{B_k}^{(2)} \Vert}
\end{eqnarray}
Based on this scheme, the received signals at $B_k$ and $E_i$ reduce to:
\begin{align}
	y_{B_k}^n &= \hv_{AB_k} \vv_{k} s_{k}^n(W_k, W_k^x) + n_{B_k}^n, \quad \forall k \in \Kc \\
	z_{E_i}^n &=  n_{E_i}^n, \quad \forall i \in \Lc
	\label{eq:beamreceived}
\end{align} 

For the environments where Line-of-Sight (LOS) propagation model is more dominant, channels between the transmitter and the receivers are more likely to be correlated, especially if the receivers are close to each other, e.g., distances between them are shorter than 19 wavelengths \cite{He:Channel} (19 wavelengths is approximately equal to 1 and 2 meters for 2.4 and 5 GHz carrier frequencies, respectively).
Let us  consider a scenario where one of Eves is near one of Bobs.
(Such a scenario can be easily observed in conference rooms, theaters, public transportation, concert halls, restaurants, stadiums etc.)
Therefore, when the information signal that is intended to the given Bob is canceled out at the given Eve using ZFBF, the same signal also becomes very weak or even canceled out at this Bob as well.
This brings about a vulnerability issue for the designs that rely on ZFBF.
In particular, for the case of correlation between $\hv_{AE_i}$ and $\hv_{AB_k} = \sqrt{D_{AB_k}} \gv_{AB_k}$ with parameter $\rho \in [0,1]$, the following equation holds:
\begin{align}
\hv_{AE_i} = \sqrt{D_{AE_i}} ( \rho \gv_{AB_k} + \sqrt{1 - \rho^2} \gv_{AE_i})
\end{align}
where $\gv_{AB_k}$ and $\gv_{AE_i}$ are independent, i.e., $\hv_{AE_i} = \sqrt{D_{AE_i}} \gv_{AE_i}$ if there is no correlation.
$\hv_{AE_i} \vv_k = 0$ due to ZFBF.
Accordingly,
\begin{align}
\sqrt{D_{AE_i}} ( \rho \gv_{AB_k} + \sqrt{1 - \rho^2} \gv_{AE_i}) \vv_k &= 0 \\
\rho \gv_{AB_k} \vv_k + \sqrt{1 - \rho^2} \gv_{AE_i} \vv_k &= 0 \\
\gv_{AB_k} \vv_k &= \dfrac{- \sqrt{1 - \rho^2}}{\rho} \gv_{AE_i} \vv_k
\end{align}
As a result, $\hv_{AB_k} \vv_k = \sqrt{D_{AB_k}} \gv_{AB_k} \vv_k = -\sqrt{D_{AB_k}} \dfrac{\sqrt{1 - \rho^2}}{\rho} \gv_{AE_i} \vv_k$.
Therefore, as $\rho$ tends to 1, $\hv_{AB_k} \vv_k \rightarrow 0 $, which means that the information signal intended to the $k$th Bob becomes very weak at this Bob.

\subsection{Cooperative FJ} 
\label{sec:FJwC} 

Using the proposed scheme detailed in the previous subsection, communication rates of Bobs are maximized after imposing the zero-forcing constraints to cancel out the information signals at Eves and unintended Bobs.
However, the antenna constraint required by the previous strategy ($N_A > L +K -1$) may not be always satisfied.
Specifically, the number of Eves may be very large such that $L +K -1 > N_A$. Moreover, even if this constraint is satisfied, having a large number of Eves may cause a very poor system performance (in terms of secrecy sum-rate or total transmit power).
Having more Eves results in more constraints, and the number of available dimensions at Alice to beamform the information signals to the intended Bobs (diversity gain) decreases.
 
In this section, we propose a strategy that requires zero-forcing constraints only for unintended Bobs. Thus, the security constraint given in (\ref{eq:secons1}) is satisfied as previously explained.
To satisfy (\ref{eq:secons2}), Alice sends TxFJ signals such that they are canceled out at Bobs by ZFBF, and their signal strength at Eves is maximized.
This way, Bobs are not affected by the TxFJ signals, and the channels of Eves become weaker.
Applying ZFBF to TxFJ signals is a well-known technique.
This concept has been studied for various single- and multi-user scenarios since the pioneering work of Goel and Negi (\cite{Goel:Guaranteeing08, Romero-Zurita:Outage12, Li:Secure14, Zhou:Securing13, Zheng:Improving13, Yang2015, Yang2015a}). We follow the same precoder design for TxFJ signals done in the aforementioned papers.
This technique only requires the constraint $N_A > K$ rather than $N_A > (L +K -1)$.

Based on the proposed scheme, the transmitted signal at Alice is given by (\ref{eq:generalsignaling}).
The precoders of the information signals are designed as follows. Let us define
\begin{align}
\hat{\Hm}_{B_k} = [ \hv_{AB_1}^T \cdots \hv_{AB_{k-1}}^T \hv_{AB_{k+1}}^T \cdots \hv_{AB_K}^T]^T, \quad \forall k \in \Kc
\end{align} 
Let the SVD of this matrix be $\hat{\Hm}_{B_k} = \hat{\Um}_{B_k} \hat{\Sigmam}_{B_k} \hat{\Vm}_{B_k}^* $.
We assume that $N_A > (K - 1)$, and $\mathrm{rank}(\hat{\Hm}_{B_k}) = U_2$. Let $\hat{\Vm}_{B_k}^{(2)}$  correspond to the last $(N_A - U_2)$ columns of $\hat{\Vm}_{B_k}$. Then, $\hat{\Vm}_{B_k}^{(2)}$ forms an orthogonal basis for the null space of $\hat{\Hm}_{B_k}$. By following the same steps as we did in the previous section, the precoders of the information signals are given as:
\begin{align}
\vv_k = \hat{\Vm}_{B_k}^{(2)} \frac{(\hv_{AB_k}\hat{\Vm}_{B_k}^{(2)})^*}{ \Vert \hv_{AB_k}\hat{\Vm}_{B_k}^{(2)} \Vert}, \quad \forall k \in \Kc
\label{eq:firstzeroforcing}
\end{align}

The precoding design of TxFJ signals is as follows. First, let us define
\begin{eqnarray}
\Hm_{AB} = [ \hv_{AB_1}^T \cdots \hv_{AB_K}^T]^T
\end{eqnarray} 
Let the SVD of this matrix be $\Hm_{AB} = \Um_{AB} \Sigmam_{AB} \Vm_{AB}^* $.
We assume that $N_A > K$, and $\mathrm{rank}(\Hm_{AB}) = U_3$.
($U_3 = K$ if $\Hm_{AB}$ is a full-rank matrix).
Let $\Vm_{AB}^{(2)}$  correspond to the last $(N_A - U_3)$ columns of $\Vm_{AB}$. Then, $\Vm_{AB}^{(2)}$ forms an orthogonal basis for the null space of $\Hm_{AB}$. As a result, each column of $\Vm_{AB}^{(2)}$ corresponds to the precoder of an independent TxFJ signal so that the null space of the channel matrix between Alice and Bobs is fully covered by TxFJ signals.
This also implies that $M = N_A - U_3$.
If $\Vm_{AB}^{(2)}(m)$ represents the $m$th column of that matrix, TxFJ signal precoders are given by: 
\begin{align}
\vv_m^{(j)} =  \Vm_{AB}^{(2)}(m), \quad \forall m \in \{1, \cdots, N_A - U_3\}
\label{eq:secondzeroforcing}
\end{align}

The precoder design of the information and TxFJ signals does not rely on the knowledge of ECSI. In the rest of the paper, the same precoders will be used in both cases where ECSI is known or unknown. Note that these precoders are unit vectors in the corresponding directions. How to allocate power for these signals in the given directions will be discussed in the next section. 

For scenarios in which LOS propogation is dominant (like the previous scenario), let us assume that one of Eves and one of Bobs are close to each other so that their channels are highly correlated.
Then, as the TxFJ signals are zero-forced at Bob, their effect becomes weak or even vanished at the given Eve as well (we discussed a similar scenario in the previous subsection).
At the same time, the information signal intended to the given Bob is sent in the direction of his channel after being zero-forced at unintended Bobs.
This is a maximum-ratio combining (MRC) precoder design, which maximizes SINR at the given receiver, with a zero-forcing constraint.
This precoder is expressed as follows:
\begin{align}
\vv_k = & \underset{  \begin{subarray} \\
 \vv \end{subarray} }{\text{argmax}} ( \hv_{AB_k} \vv ) =
 \underset{  \begin{subarray} \\
 \vv \end{subarray} }{\text{argmax}} ( \gv_{AB_k} \vv ) \\
 & \quad s.t. \quad \hat{\Hm}_{B_k} \vv = 0
\end{align}
In the case of channel correlation between $\hv_{AE_i}$ and $\hv_{AB_k}$, $\hv_{AE_i} = \sqrt{D_{AE_i}} ( \rho \gv_{AB_k} + \sqrt{1 - \rho^2} \gv_{AE_i})$ as discussed earlier.
The multiplication of $\hv_{AE_i}$ and $\vv_k$ becomes:
\begin{align}
\hv_{AE_i} \vv_k = \sqrt{D_{AE_i}} \rho \gv_{AB_k} \vv_k + \sqrt{D_{AE_i}} \sqrt{1 - \rho^2} \gv_{AE_i} \vv_k
\end{align}
$\vv_k$ maximizes the first term of $\hv_{AE_i}$.
However, if $\rho$ is low, this term will be negligible.
On the other hand, if $\rho$ is significantly larger than  $\sqrt{1 - \rho^2}$, the first term will be dominant.
Due to these two reasons, SINR at Eve increases in the case of high channel correlation.
To overcome this problem, we utilize FD communications.
In our model, Bobs are capable of transmitting and receiving signals over the same frequency band at the same time.
As a result, we propose sending RxFJ signals from Bobs.
That is, while TxFJ ensures that Eves (whose channels are uncorrelated with Bobs) are jammed, RxFJ aims to keep the vicinity of Bobs secure.
Besides, whenever a new Bob is served by Alice, one TxFJ dimension is sacrificed.
However, the total number of dimensions occupied by TxFJ and RxFJ remains the same, when this Bob generates his own RxFJ.
This is an important point, as more dimensions allow to design more effective friendly jamming signals.

\section{Optimal Power Allocation}   
\label{sec:SQoS}

\subsection{Known ECSI}
In this section, we consider a problem that aims to minimize the total power allocated to the information, TxFJ, and RxFJ signals while maintaining certain secrecy rate requirements.
These requirements ensure that the mutual information between the information signal, $S_k$, and the received signal at the intended receiver, $Y_{B_k}$, is above a certain threshold, $R_k + R_k^x$ (sum of the individual secrecy rate and the randomization rate of $s_k$) $\forall k \in \Kc$, and the mutual information between the information signal, $S_k$, and the received signal at the $i$th Eve, $Z_{E_i}$, is below a certain threshold, $R_k^x$ (the randomization rate of $s_k$) $\forall k \in \Kc$ and $\forall i \in \Lc$.
Furthermore, we assume that the power constraints given in (\ref{eq:powerconsttt}) and (\ref{eq:powerconst}) still need to be satisfied.

Here, we assume that Alice knows the channels between herself and all the receivers including Eves, and the channels between each receiver pair (including the channels between Bobs and Eves).
This assumption holds for a network where Alice is a BS, and Bobs and Eves are active and idle system devices, respectively.
IEEE 802.11ac is a well-suited standard for this system model, as it allows multiuser downlink transmission through beamforming.
An instance of our setup consists of $K+L$ legitimate receivers, where Alice is capable of serving only $K$ of them simultaneously.
For example, the maximum number of concurrent transmissions in 802.11ac is 4 (i.e., $K \leq 4$).
At the beginning of each transmission block (or, coherence interval), Alice can acquire the CSI of all $K+L$ receivers to decide which $K$ of them will be served.
The channel estimation is performed via \textit{explicit} or \textit{implicit beamforming} in 802.11ac systems.
Explicit beamforming relies on packet exchanges between Alice and receivers.
Specifically, Alice transmits an NDP (Null Data Packet) following the NDP announcement message.
Then, the first receiver sends its feedback to Alice, providing its estimate of the CSI.
After that, Alice polls the other receivers successively, and they send their feedback to Alice similarly.
This way, the CSI between Alice and receivers is extracted.
Alternatively, Alice can estimate the CSI of any receiver based on known fields (e.g., preambles) of its current transmissions.
This method is called implicit beamforming, and relies on channel reciprocity.
After CSI acquisition, each of the $K$ receivers (Bobs) that are selected to be served receives a message that should be kept confidential from the other $K+L-1$ receivers.
Therefore, even though the receivers in this system model are not necessarily malicious, they are ``untrusted".
For instance, these receivers may be compromised or hacked by an external attacker.
Information-theoretic security of these $K$ information messages is guaranteed by zero-forcing precoding against $K-1$ Bobs.
The remaining $L$ idle receivers that are not selected to be served are treated as eavesdroppers but with known CSI.

Another instance of our setup involves multiple adjacent 802.11ac networks.
Users belonging to any adjacent network can be treated as external eavesdroppers from the standpoint of the given network (again, they may be compromised).
The packets sent by these users include known 802.11ac headers, and may be overheard by Alice.
Accordingly, Alice can estimate the CSI between her and these adjacent users using implicit beamforming.

Likewise, the CSI between each receiver pair (Bob-Bob) and eavesdropper-receiver pair (Eve-Bob) can be estimated through \textit{implicit beamforming}.
For example, any transmitted message from Bob and Eve includes short- and long-training sequences (which are a part of the known preamble) that facilitate channel estimation.
Therefore, any Bob can estimate the channels from the other Bobs and between Eves and the given Bob by overhearing and processing these messages.
Alice can then acquire the estimated CSI by polling each Bob.

Consequently, the problem formulation is given by:
\begin{subequations}
\label{prob:1}
\begin{align}
\underset{ 
\begin{subarray} \\
P_{S_k} \; \forall k \in \Kc \\
P_m^{(j)} \; \forall m \in \Mc \\
P_{B_k} \; \forall k \in \Kc \end{subarray} }{\text{minimize}} 
& \sum_{k \in \Kc} P_{S_k} +  \sum_{m \in \Mc} P_{m}^{(j)} + \sum_{k \in \Kc} P_{B_k} \\
 s.t. \quad \quad & \sum_{k \in \Kc} P_{S_k} +  \sum_{m \in \Mc} P_m^{(j)} \leq \bar{P}_{A}  \\
& P_{B_k} \leq \bar{P}_{B_k}, \; \forall k \in \Kc \\
& I ( S_{k}; Y_{B_k}) \geq R_k + R_k^x, \; \forall k \in \Kc  \label{const:1} \\
& I ( S_k; Z_{E_i} ) \leq R_k^x, \; \forall k \in \Kc, \; \forall i \in \Lc
\label{const:2}
\end{align}
\end{subequations}
where $\Kc = \{1, \cdots, K\}$, $\Mc = \{1 , \cdots, N_A - U_3\}$, and $ \Lc = \{ 1, \cdots, L\}$. Given the communication scheme described in the previous section, the mutual information between $S_k$ and $Y_{B_k}$ is given by:
\begin{align}
I ( S_{k}; Y_{B_k})= \log ( 1 + \mathrm{SINR}_{B_k} ), \; \forall k \in \Kc
\label{eq:sinrbob}
\end{align}	
where
$$ \mathrm{SINR}_{B_k} = \dfrac{ P_{S_k} \vert \hv_{AB_k}\vv_{k} \vert^{2} }{ \alpha P_{B_k} \vert h_{B_kB_k} \vert^{2} + \sum_{l \in \{ \Kc \setminus k \}} P_{B_l} \vert h_{B_lB_k} \vert^{2} + N_0 }.$$
Similarly, the mutual information between $S_k$ and $Z_{E_i}$ is given by:
\begin{align} \label{eqn:sinreve}
I ( S_k; Z_{E_i} ) = \log ( 1 + \dfrac{  P_{S_k} \vert \hv_{AE_i}\vv_{k} \vert^{2} }{  A + B + C + N_0 } ), \quad \forall k \in \Kc, \; \forall i \in \Lc 
\end{align}
where $A = \sum_{l \in \{ \Kc \setminus k \}} P_{S_l} \vert \hv_{AE_i}\vv_l \vert^2$, $ B = \sum_{m \in \Mc} P_{m}^{(j)} \vert \hv_{AE_i} \vv_m^{(j)} \vert^{2}$, and $C = \sum_{k \in \Kc} P_{B_k} \vert h_{B_kE_i} \vert^{2}$. $A$, $B$, and $C$ are the interference terms due to other information, TxFJ, and RxFJ signals, respectively.
Note that the interfering information signals help each other by decreasing the SINR at each Eve, as we consider the individual secrecy rates.
Based on (\ref{eq:sinrbob}) and (\ref{eqn:sinreve}), the constraints in (\ref{const:1}) and (\ref{const:2}) are given by:
\begin{align}
P_{S_k} \vert \hv_{AB_k}\vv_{k} \vert^{2} & \geq (2^{R_k + R_k^x} -1)(\alpha P_{B_k} \vert h_{B_kB_k} \vert^{2} + \sum_{l \in \{\Kc \setminus k \}} P_{B_l} \vert h_{B_lB_k} \vert^{2} + N_0), \; \forall k \in \Kc \\
P_{S_k} \vert \hv_{AE_i}\vv_{k} \vert^{2} & \leq (2^{R_k^x} -1)( A + B + C + N_0), \; \forall k \in \Kc, \; \forall i \in \Lc
\end{align}
As a result, we have a linear programming problem, as all of the constraints and the objective function are linear. 
The achievable individual secrecy rate for $B_k$ satisfies the following inequality:
\begin{align}
R_{k} \leq [I ( S_{k}; Y_{B_k}) - I ( S_{k}; Z_{E_i})]^+, \quad \forall i \in \Lc 
\end{align}
Therefore, instead of separate secrecy requirements as in (\ref{const:1}) and (\ref{const:2}), Bobs may request a certain individual secrecy rate.
In particular, the constraints in (\ref{const:1}) and (\ref{const:2}) can be replaced as follows:
\begin{subequations}
\label{eq:realproblem}
\begin{align}
\underset{ 
\begin{subarray} \\
P_{S_k} \; \forall k \in \Kc \\
P_m^{(j)} \; \forall m \in \Mc \\
P_{B_k} \; \forall k \in \Kc \end{subarray} }{\text{minimize}} 
& \sum_{k \in \Kc} P_{S_k} +  \sum_{m \in \Mc} P_{m}^{(j)} + \sum_{k \in \Kc} P_{B_k} \\
 s.t. \quad \quad & \sum_{k \in \Kc} P_{S_k} +  \sum_{m \in \Mc} P_m^{(j)} \leq \bar{P}_{A}  \\
& P_{B_k} \leq \bar{P}_{B_k}, \; \forall k \in \Kc \\
& I ( S_{k}; Y_{B_k}) - I ( S_{k}; Z_{E_i}) \geq R_k , \; \forall k \in \Kc, \forall i \in \Lc 
\end{align}
\end{subequations}
where $R_k$ is a nonnegative individual secrecy rate. However, this makes the problem non-convex. Here, the same problem formulation in (\ref{prob:1}) can be used for a given set of randomization rate values $R_k^x$. Note that, $R_k^x$ is ``the designed randomization rate" that confuses Eves, and the problem reduces to choosing optimal amount of randomization to minimize the total power consumption while satisfying the individual secrecy rate requirements. It can be found by a line search method.

Note that even if the CSI of Bobs and the corresponding Eves is correlated, the same analysis holds.
The other issue is how to ensure that Bobs generate the RxFJ signals as designed at Alice, as they are not trustworthy devices.
The objective of sending an RxFJ signal from each Bob is to provide security only around that Bob (i.e., the power of the RxFJ signal is limited by design).
Therefore, Bobs are not supposed to help each other.
Even if Bobs rely on each other to degrade Eves, Alice computes how much power Bobs should allocate to their RxFJ signals, as she is the only one who has all the necessary parameters (e.g., CSI) to solve the optimization problem. 
A given Bob does not exactly know whether his RxFJ can harm other Bobs.
Even if some of Bobs behave in an adversarial manner by using their RxFJ signal for malicious (jamming) purposes, they can be easily detected by Alice, and  prevented from transmitting (note that such active-attack model is not part of our underlying setup).

\subsection{Unknown ECSI}

In this section, we assume that the first- and second-order statistics of the ECSI are known, not the ECSI itself.
How Alice obtains the perfect CSI between Bobs and herself in an 802.11ac network is explained in the previous section.
However, after the last acquisition of perfect CSI for a given Eve, she may move to another location or the small-scale effects of her channel may change.
As a result, some perturbation relative to the last known CSI of this Eve can be assumed.
Based on this information, Alice can estimate Eve's channel statistics.
Specifically, we assume that Alice knows
\begin{align}
\Km_{AE_i} & = \mathbb{E}[\hv_{AE_i}^*\hv_{AE_i}] \\
\mu_{B_kE_i} & = \mathbb{E}[h_{B_kE_i}^*h_{B_kE_i}]
\end{align}
$\forall i \in \Lc$ and $\forall k \in \Kc$.
We consider replacing the randomization rate constraint in (\ref{const:2}) with an outage constraint for all Bobs, as ECSI is random.
The probability of having at least one Eve such that the mutual information between the received signal at this Eve and the information signal, $S_k$, is greater than or equal to the designed randomization rate, $R_k^x$, is called the outage probability for the $k$th Bob.
The outage constraint states that the outage probability should be smaller than or equal to a certain constant $\epsilon_k$ for the $k$th Bob.
Particularly, if there exists only one Eve, i.e. $L=1$, this constraint is given by:
\begin{align}
\Pr \{I ( S_{k}; Z_{E_1}) \geq R_k^x \} \leq \epsilon_k, \; \forall k \in \Kc 
\label{eq:initialoutage}
\end{align}
In the presence of $L$ non-colluding Eves, this outage probability becomes:
\begin{align}
1 - ( 1 - \Pr \{I ( S_{k}; Z_{E_i}) \geq R_k^x \} )^L & \leq \epsilon_k, \; \forall k \in \Kc, \; i \in \Lc   \label{const:outage0}\\
\Pr \{ I ( S_{k}; Z_{E_i}) \geq R_k^x \} & \leq 1 - \sqrt[L]{1 - \epsilon_k} , \; \forall k \in \Kc, \; i \in \Lc
\label{const:outage1}
\end{align}
where we assume all Eves have the same channel properties, and the channels between Alice and each Eve are independent.
(The inequality in (\ref{const:outage1}) is identical to the one in (\ref{eq:initialoutage}) when $L=1$.)
Note that this independence assumption brings about the worst-case scenario.
(Otherwise, the right hand side of the inequality in (\ref{const:outage1}) would take a larger value, so satisfying the outage probability constraint would be easier.)
Therefore, it does not contradict the assumption that the CSI of Bobs and the corresponding Eves is correlated.
By integrating the equation in (\ref{eqn:sinreve}) into this outage probability expression, we obtain the first and the second equalities in (\ref{eq:lastoutage}) where $D = \sum_{l \in \{ \Kc \setminus k \}} P_{S_l} \vv_{l}^* \hv_{AE_i}^* \hv_{AE_i} \vv_{l}$, $F = \sum_{m \in \Mc} P_{m}^{(j)}  (\vv_m^{(j)})^* \hv_{AE_i}^* \hv_{AE_i} \vv_m^{(j)}$, and $G = \sum_{k \in \Kc} P_{B_k} h_{B_kE_i}^* h_{B_kE_i}$. Nevertheless, it is not possible to obtain a tractable problem by using this outage constraint. Thus, we exploit Markov's inequality, which states the following:
\begin{align}
\Pr \{ X \geq a \} \leq \dfrac{\mathbb{E}[X]}{a}
\end{align}
where $a > \mathbb{E}[X]$. Therefore, the outage expression can be upper-bounded using Markov's inequality as in the third expression in (\ref{eq:lastoutage}). By assuming the channels are zero mean, this is modified as in the forth expression, where $\bar{D} = \sum_{l \in \{\Kc \setminus k\}} P_{S_l} \vv_{l}^* \Km_{AE_i} \vv_{l}$, $\bar{F} =  \sum_{m \in \Mc} P_{m}^{(j)}  (\vv_m^{(j)})^* \Km_{AE_i} \vv_m^{(j)}$, and $\bar{G} = \sum_{k \in \Kc} P_{B_k} \mu_{B_kE_i}$. Note that a similar inequality can be written for channels with non-zero mean. As a result, the constraint (\ref{const:outage1}) is converted to the constraint in the last equation of (\ref{eq:lastoutage}) for all $ k \in \Kc$.
(The upper bound obtained by Markov's inequality is used for outage probability, so the analysis here is on the conservative side. One can utilize tighter bounds like Chebyshev's or Chernoff's inequalities, but we do not pursue this here. We note that a similar Markov bound was used in \cite{gerbracht2012secrecy}.)

\begin{figure*}[!t]
\normalsize
\setcounter{equation}{38}
\begin{align*}
\Pr \{I ( S_{k}; Z_{E_i}) \geq R_k^x \} & = \Pr \{\log ( 1 + \dfrac{  P_{S_k} \vv_{k}^* \hv_{AE_i}^* \hv_{AE_i} \vv_{k} }{ D + F + G + 1 } ) \geq R_k^x \} \\
& = \Pr \{ P_{S_k} \vv_{k}^* \hv_{AE_i}^* \hv_{AE_i} \vv_{k} - (2^{R_k^x} -1)( D + F + G) \geq 2^{R_k^x} -1 \} \\
& \leq \dfrac{\mathbb{E}[P_{S_k} \vv_{k}^* \hv_{AE_i}^* \hv_{AE_i} \vv_{k} - (2^{R_k^x} -1)( D + F + G)]}{2^{R_k^x} -1} \\
& = \dfrac{P_{S_k} \vv_{k}^* \Km_{AE_i} \vv_{k} - (2^{R_k^x} -1)( \bar{D} + \bar{F} + \bar{G})}{2^{R_k^x} -1}
\end{align*}
\begin{align}
\dfrac{P_{S_k} \vv_{k}^* \Km_{AE_i} \vv_{k} - (2^{R_k^x} -1)( \bar{D} + \bar{F} + \bar{G})}{2^{R_k^x} -1} & \leq 1 - \sqrt[L]{1 - \epsilon_k}
\label{eq:lastoutage}
\end{align}
\hrulefill
\vspace*{4pt}
\end{figure*}
If the CSI of the $k$th Bob and the $i$th Eve is correlated with parameter $\rho_{ki}$, the analysis is modified as follows.
The relationship between $\hv_{AE_i}$ and $\hv_{AB_k} = \sqrt{D_{AB_k}} \gv_{AB_k}$ becomes:
\begin{align}
\hv_{AE_i} = \sqrt{D_{AE_i}} ( \rho_{ki} \gv_{AB_k} + \sqrt{1 - \rho_{ki}^2} \gv_{AE_i} )
\end{align}
where $\gv_{AE_i}$ and $\gv_{AB_k}$ are independent of each other, i.e., $\hv_{AE_i} = \sqrt{D_{AE_i}} \gv_{AE_i}$ when $\rho_{ki} = 0$.
Therefore, the covariance matrix $\Km_{AE_i}$ is formed by:
\begin{align}
\Km_{AE_i} &=  \mathbb{E}[\hv_{AE_i}^*\hv_{AE_i}] \\
&= \mathbb{E}[ D_{AE_i} (\rho_{ki} \gv_{AB_k} + \sqrt{1 - \rho_{ki}^2} \gv_{AE_i})^* (\rho_{ki} \gv_{AB_k} + \sqrt{1 - \rho_{ki}^2} \gv_{AE_i}) ] \\
&= \mathbb{E}[ D_{AE_i} \rho_{ki}^2 \gv_{AB_k}^* \gv_{AB_k} ] + \mathbb{E}[ D_{AE_i} (1  - \rho_{ki}^2) \gv_{AE_i}^* \gv_{AE_i} ] + \mathbb{E} [ 2 D_{AE_i} \rho_{ki} \sqrt{ 1 - \rho_{ki}^2} \; \mathbb{R} \{ \gv_{AB_k}^* \gv_{AE_i} \} ] \\
&=  \rho_{ki}^2 \gv_{AB_k}^* \gv_{AB_k} \mathbb{E}[ D_{AE_i} ] + (1  - \rho_{ki}^2) \mathbb{E}[ D_{AE_i} ]\mathbb{E}[  \gv_{AE_i}^* \gv_{AE_i} ] + 2 \rho_{ki} \sqrt{ 1 - \rho_{ki}^2} \mathbb{E}[ D_{AE_i} ] \; \mathbb{R} \{ \hv_{AB_k}^* \mathbb{E} [ \gv_{AE_i} ] \}  \label{eq:b} \\
&=  \rho_{ki}^2 \gv_{AB_k}^* \gv_{AB_k} \mathbb{E}[ D_{AE_i} ] + (1  - \rho_{ki}^2) \mathbb{E}[ D_{AE_i} ] \textbf{I}_{N_A} \label{eq:a}
\end{align}
where $\mathbb{R} \{ . \}$ represents the real part of a complex number.
The equation (\ref{eq:b}) follows that $\gv_{AB_k}$ is a known vector, which is small-scale channel effects between Alice and the $k$th Bob, and $ D_{AE_i}$ and $\gv_{AE_i}$ are independent random variables.
Furthermore, the last equality follows that $\mathbb{E} [ \gv_{AE_i} ] = 0$ and $\mathbb{E}[  \gv_{AE_i}^* \gv_{AE_i} ] = \textbf{I}_{N_A}$, as $ \gv_{AE_i} \sim \Cc \Nc (0, \textbf{I}_{N_A})$.
The first- and second-order statistics of $D_{AE_i}$ are known as explained before, so $\Km_{AE_i}$ can be estimated by (\ref{eq:a}).
Obtaining the exact correlation coefficient is not possible, if ECSI is unknown.
Therefore, this correlation coefficient can be treated as a controllable security metric.
For example, Bobs may request a certain correlation coefficient based on the secrecy level they would like to achieve (e.g., if Bobs assume that there is another device nearby, they request a higher correlation coefficient).
Similarly, Alice may guarantee secure communication for Bobs only up to a certain level of correlation for a given $\epsilon_k$.
Note that there is a tradeoff between correlation coefficient and outage probability.
If Bobs request Alice to use a larger $\rho_{ki}$ to increase the secrecy level, it will be harder to satisfy the outage probability.
In this case, Alice may not find a feasible solution, and she needs to increase $\epsilon_k$ to relax the outage probability constraint.
As we previously assume that all Eves have the same channel properties, the second subscript of the correlation coefficient can be omitted, i.e., $\rho_{ki} = \rho_k$ $\forall i \in  \Lc$.
Consequently, the outage probability constraint is given by (\ref{eq:lastoutage}) where $\Km_{AE_i}$ is calculated using (\ref{eq:a}).

\section{Simulation Results and Discussions}

We simulate an 802.11ac network in a simplified manner, using some of its system parameters.
The coherence time of the channels is large enough so that Alice perfectly acquires the CSI of Bobs and Eves via previously explained channel estimation techniques.
Therefore, within each coherence interval (or transmission block), the channels are constant, whereas at the beginning of each coherence interval, the channels from $i \in \{ A \cup \Bc \}$ to $j \in \{ \Bc \cup \Ec \}$ are randomly generated as
\begin{align}
\hv_{ij} = \gv_{ij} \sqrt{ A (\frac{c}{4 \pi f d_{ij} })^{3} }  
\end{align}
where $A$, $c$, $f$, and $d_{ij}$ are the antenna gain, speed of light, operating center frequency, and distance between the corresponding devices, respectively.
($A$ is assumed to be the same between each $i$-$j$ pair.)
This is a modified version of the simplest form of Friis transmission equation.
$\gv_{ij} \sim \Cc\Nc(0, \textbf{I}_{N_A})$ represents small-scale effects of the channel.
We set $A$ to $4$ and $f$ to $5260$ MHz throughout the simulations, while the frequency bandwidth is $160$ MHz.
The maximum power outputs at Alice, $\bar{P}_A$, and each receiver, $\bar{P}_{B_k}$ $\forall k \in \Kc$, are $24$ dBm and $10$ dBm, respectively. 
The thermal noise for $160$ MHz bandwidth is $-95$ dBm.
The number of antennas at Alice, $N_A$, is set to $8$, and she can serve at most $4$ Bobs simultaneously.
We assume that Bobs and Eves are uniformly and randomly distributed in a circular area around Alice with a radius of $30$ meters unless otherwise stated.
We show the average value of $2000$ different realizations of the network in the simulation results.

\subsection{Known ECSI}

\begin{figure}[!t]
    \centering
    \begin{subfigure}[b]{0.475\textwidth}
        \centering
        \includegraphics[width=90mm]{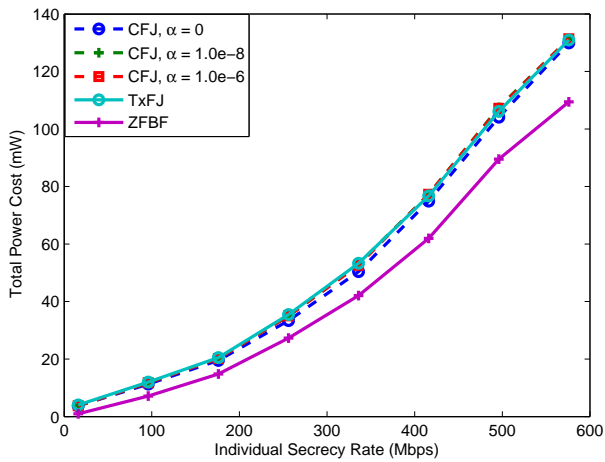}
        \caption{Independently located Bobs and Eves}
        \label{fig:2a}
    \end{subfigure}%
    \hfill
    \begin{subfigure}[b]{0.475\textwidth}
        \centering
        \includegraphics[width=90mm]{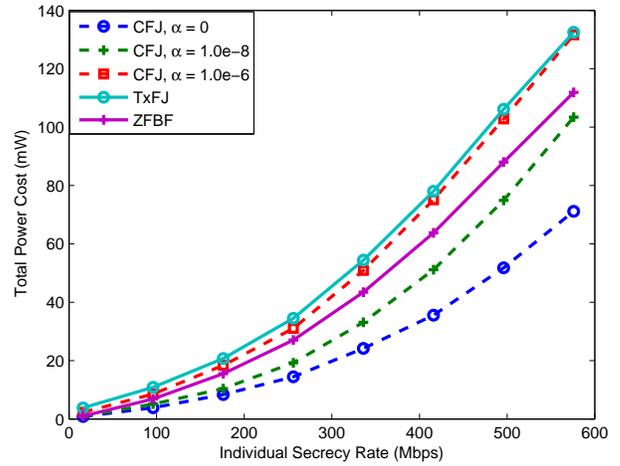}
        \caption{Independently located Bobs with nearby Eves ($\rho = 0$)}
        \label{fig:2b}
    \end{subfigure}
    \vskip\baselineskip
    \begin{subfigure}[b]{0.475\textwidth}
        \centering
        \includegraphics[width=90mm]{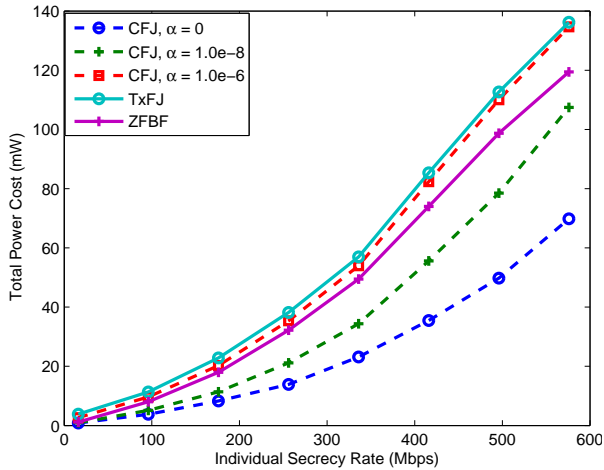}
        \caption{Independently located Bobs with nearby Eves ($\rho = 0.4$)}
        \label{fig:2c}
    \end{subfigure}
    \quad
    \begin{subfigure}[b]{0.475\textwidth}
        \centering
        \includegraphics[width=90mm]{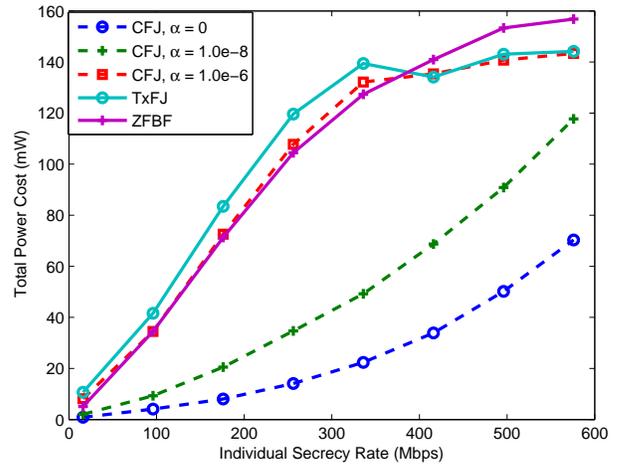}
        \caption{Independently located Bobs with nearby Eves ($\rho = 0.9$)}
        \label{fig:2d}
    \end{subfigure}    
    \caption{Total power consumption vs. individual secrecy rate with known ECSI where $K = 3$ and $L = 3$.}
    \label{fig:2}
\end{figure}
ZFBF and cooperative FJ (CFJ) techniques are introduced in Sections \ref{sec:ZF} and \ref{sec:FJwC}, respectively.
Fig. \ref{fig:2} shows the performance evaluation of these techniques and TxFJ (without using RxFJ) with known ECSI in terms of power consumption for the given individual secrecy rates.
The problem given in (\ref{eq:realproblem}) is solved, when all of Bobs demand the same individual secrecy rate, i.e. $R_k = R_l$ $\forall k,l \in \Kc$.
(This problem is formulated for CFJ. However, the same formulation can be used to solve TxFJ problem as well, when Bobs have no power for RxFJ.) 
3 Bobs and 3 Eves are assumed to be present.
In Fig. \ref{fig:2a}, they are randomly and uniformly located around Alice as specified before.
In this case, there is no correlation between the channels of Bobs and Eves.
In Figs. \ref{fig:2b}, \ref{fig:2c}, and \ref{fig:2d}, Bobs are located similarly.
However, each Eve randomly selects one of Bobs, and she is randomly and uniformly located in a circular area around him with a radius of $1$ meter (no closer than $10$ cm). 
The correlations between the channels of Bobs and the corresponding Eves are $0$, $0.4$, and $0.9$ in Figs. \ref{fig:2b}, \ref{fig:2c}, and \ref{fig:2d}, respectively.
Furthermore, to observe the effect of SIS, three different values of $\alpha$ are considered as follows.
The self-interference channel is modeled as $\hv_{B_kB_k} = 1$ $\forall k \in \Kc$. 
If Bobs use all of their powers, the self-interference becomes $10$ dBm without any suppression. 
When $\alpha = 0$, the self-interference is assumed to be negligible compared to the noise floor, which is $-95$ dBm.
When $\alpha$ is equal to $1.0e-8$ or $1.0e-6$, the self-interference becomes $-70$ dBm (corresponding to $80$ dB suppression) or $-50$ dBm (corresponding to $60$ dB suppression), respectively.
(Note that $60$ dB suppression can be easily achieved employing the full-duplex radio design techniques in the literature \cite{Bharadia:Full13, duarte2012experiment, balatsoukas2013self, knox2012single}.) 

Fig. \ref{fig:2a} shows that ZFBF outperforms the other schemes for the given setup, when Bobs and Eves are independently located.
Moreover, the performances of TxFJ and CFJ are identical, which means that RxFJ is not employed in this case (it is not optimal).
The performance of TxFJ and ZFBF does not change, when Bobs have nearby Eves as in Fig. \ref{fig:2b}, since the channels of Bobs and Eves are still independent.
However, CFJ with high SIS starts outperforming TxFJ and ZFBF.
Employing RxFJ becomes optimal, as Eves are closer to Bobs.
When $\rho = 0.4$ as in Fig. \ref{fig:2c}, power consumption of all schemes slightly increases.
The performance gain of CFJ (with high SIS) relative to others also increases.
In Fig. \ref{fig:2d}, a high channel correlation case is investigated. 
The performance of CFJ with high SIS is much better than TxFJ and ZFBF.
Indeed, power consumption of CFJ does not change much with $\rho$, as a small amount of power for RxFJ signals is adequate to satisfy the individual secrecy rate constraints.
On the other hand, TxFJ and ZFBF need to spend much more power in high channel correlation case.
(There is a discrepancy in Fig. \ref{fig:2d}, as there is no feasible solution for TxFJ most of the time.)

\subsection{Unknown ECSI}

\begin{figure}[!t]
    \centering
    \begin{subfigure}[b]{0.475\textwidth}
        \centering
        \includegraphics[width=90mm]{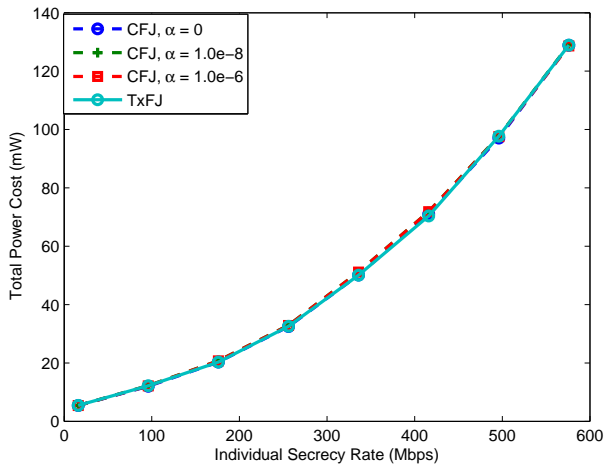}
        \caption{Independently located Bobs and Eves}
        \label{fig:3a}
    \end{subfigure}%
    \hfill
    \begin{subfigure}[b]{0.475\textwidth}
        \centering
        \includegraphics[width=90mm]{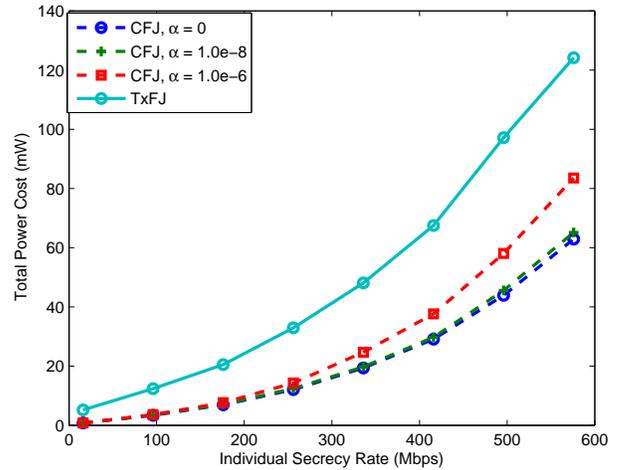}
        \caption{Independently located Bobs with nearby Eves ($\rho = 0$)}
        \label{fig:3b}
    \end{subfigure}
    \vskip\baselineskip
    \begin{subfigure}[b]{0.475\textwidth}
        \centering
        \includegraphics[width=90mm]{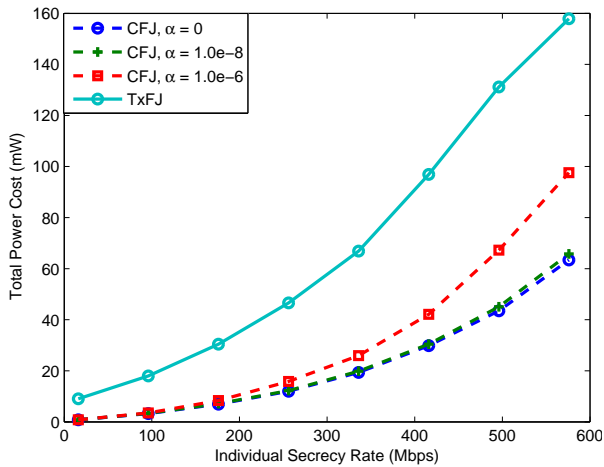}
        \caption{Independently located Bobs with nearby Eves ($\rho = 0.4$)}
        \label{fig:3c}
    \end{subfigure}
    \quad
    \begin{subfigure}[b]{0.475\textwidth}
        \centering
        \includegraphics[width=90mm]{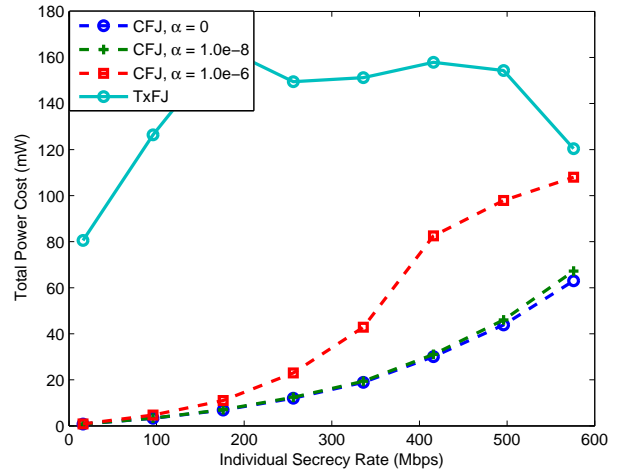}
        \caption{Independently located Bobs with nearby Eves ($\rho = 0.9$)}
        \label{fig:3d}
    \end{subfigure}    
    \caption{Total power consumption vs. individual secrecy rate with unknown ECSI where $K = 3$, $L = 3$, and $\epsilon = 0.01$.}
    \label{fig:3}
\end{figure}
In Fig. \ref{fig:3}, we show the performance of using CFJ in the scenarios where ECSI is unknown. 
As ZFBF cannot be used without the knowledge of the channels, CFJ with different levels of SIS and TxFJ are compared with each others.
The same system setup given in the previous section is used, while $\epsilon_k = 0.01$ $\forall
k \in \Kc$.
(Note that the individual secrecy outage probability should be less than or equal to $\epsilon_k$).
Again, when Bobs and Eves are independently located, CFJ and TxFJ have the same performance (RxFJ is not employed).
However, when Bobs have Eves in their vicinity, CFJ outperforms TxFJ.
While the correlation of the channels between Bobs and the corresponding Eves is increasing, the performance gain of employing RxFJ also increases.
Unlike the known ECSI case, when $\alpha = 1.0e-8$, CFJ is still much better than TxFJ.
(Again, the discrepancy in Fig. \ref{fig:3d} is due to obtaining unfeasible solutions most of the time in TxFJ scheme.)

\subsection{User Scheduling}

So far, all of $K$ Bobs were served without considering any scheduling scheme so that they could achieve the given individual secrecy rate requirements instantaneously, i.e., at each transmission time.
Here, we investigate whether the total power consumption can be reduced further by serving a subset of Bobs, while achieving the individual secrecy rate requirements of each Bob on average.
Let the probability that Alice serves the $k$th Bob in the given transmission block be $p_k$.
Accordingly, the $k$th Bob requires $p_k^{-1}$ times more individual secrecy rate per transmission block than the one in the previous sections, when he is served.
We propose to select the closest $T$ Bobs to Alice in this paper, while $\Kc_{T}$ denotes the set of indices belonging to the selected $T$ Bobs.
The rest of Bobs are treated as Eves with known CSI, i.e., the total number of Eves becomes $L + K - T$.
The problem formulation in (\ref{eq:realproblem}) is modified  as follows:
\begin{subequations}
\begin{align}
\underset{ 
\begin{subarray} \\
P_{S_k} \; \forall k \in \Kc_{T} \\
P_m^{(j)} \; \forall m \in \Mc \\
P_{B_k} \; \forall k \in \Kc_{T} \end{subarray} }{\text{minimize}} 
& \sum_{k \in \Kc_{T}} P_{S_k} +  \sum_{m \in \Mc} P_{m}^{(j)} + \sum_{k \in \Kc_{T}} P_{B_k} \\
 s.t. \quad \quad & \sum_{k \in \Kc_{T}} P_{S_k} +  \sum_{m \in \Mc} P_m^{(j)} \leq \bar{P}_{A}  \\
& P_{B_k} \leq \bar{P}_{B_k}, \; \forall k \in \Kc_{T} \\
& I ( S_{k}; Y_{B_k}) - I ( S_{k}; Z_{E_i}) \geq p_k^{-1} R_k , \; \forall k \in \Kc_{T}, \forall i \in \{ \Lc \cup \Kc \setminus \Kc_{T} \}
\end{align}
\end{subequations}
In the performance evaluation, we average the results over locations of Bobs and Eves.
Locations are constant for a transmission block, and they are randomly and uniformly chosen between blocks.
(We didn't incorporate a mobility scheme that models a more realistic network model, as this is not the scope of this paper.
However, it can be thought as Bobs are moving very fast so that the topology completely and independently changes at each block.)
Therefore, $p_k = T/K$ $\forall k \in \Kc$.
\begin{figure}[!t]
    \centering
    \begin{subfigure}[b]{0.475\textwidth}
        \centering
        \includegraphics[width=90mm]{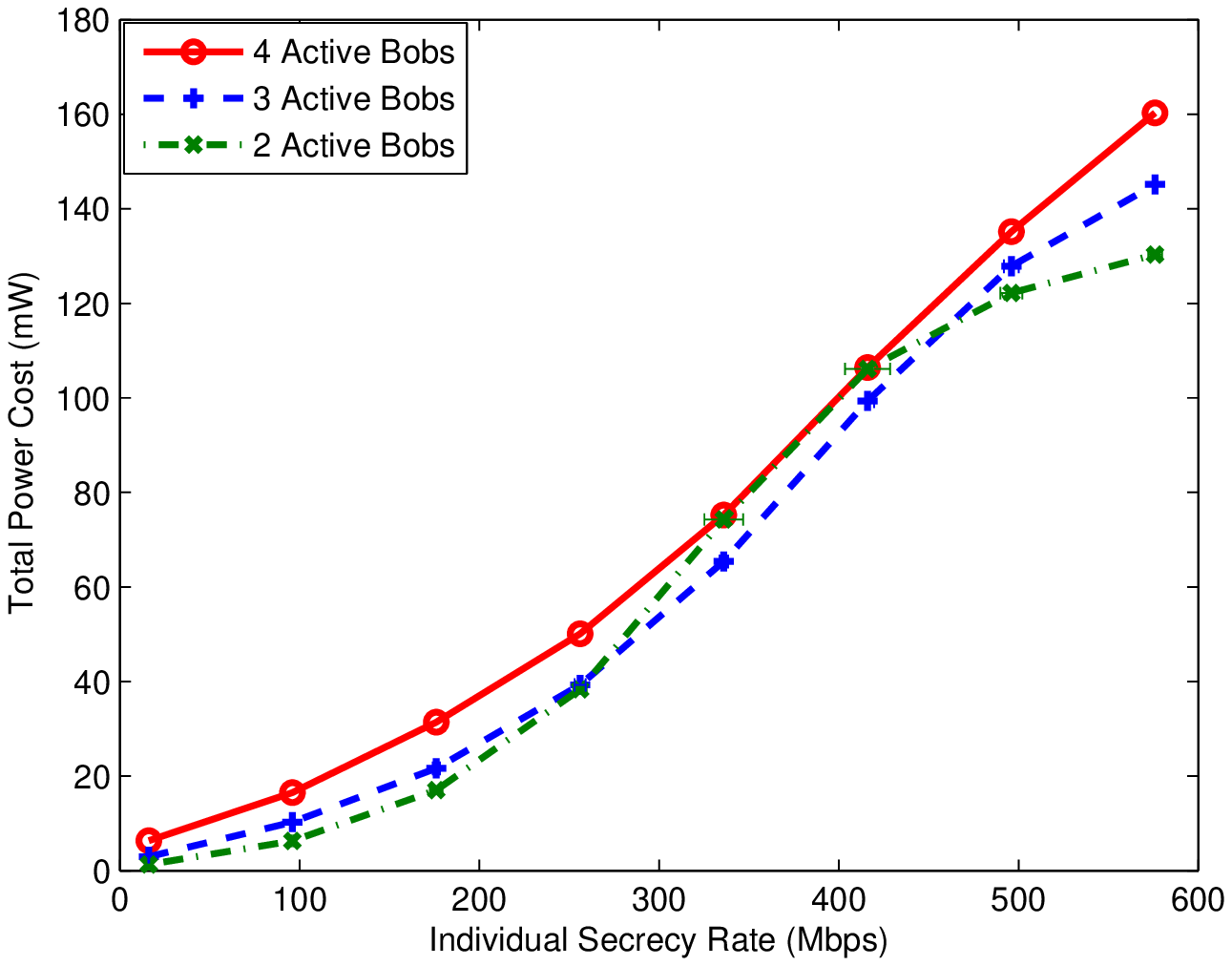}
        \caption{Independently located Bobs and Eves}
        \label{fig:4a}
    \end{subfigure}%
    \hfill
    \begin{subfigure}[b]{0.475\textwidth}
        \centering
        \includegraphics[width=90mm]{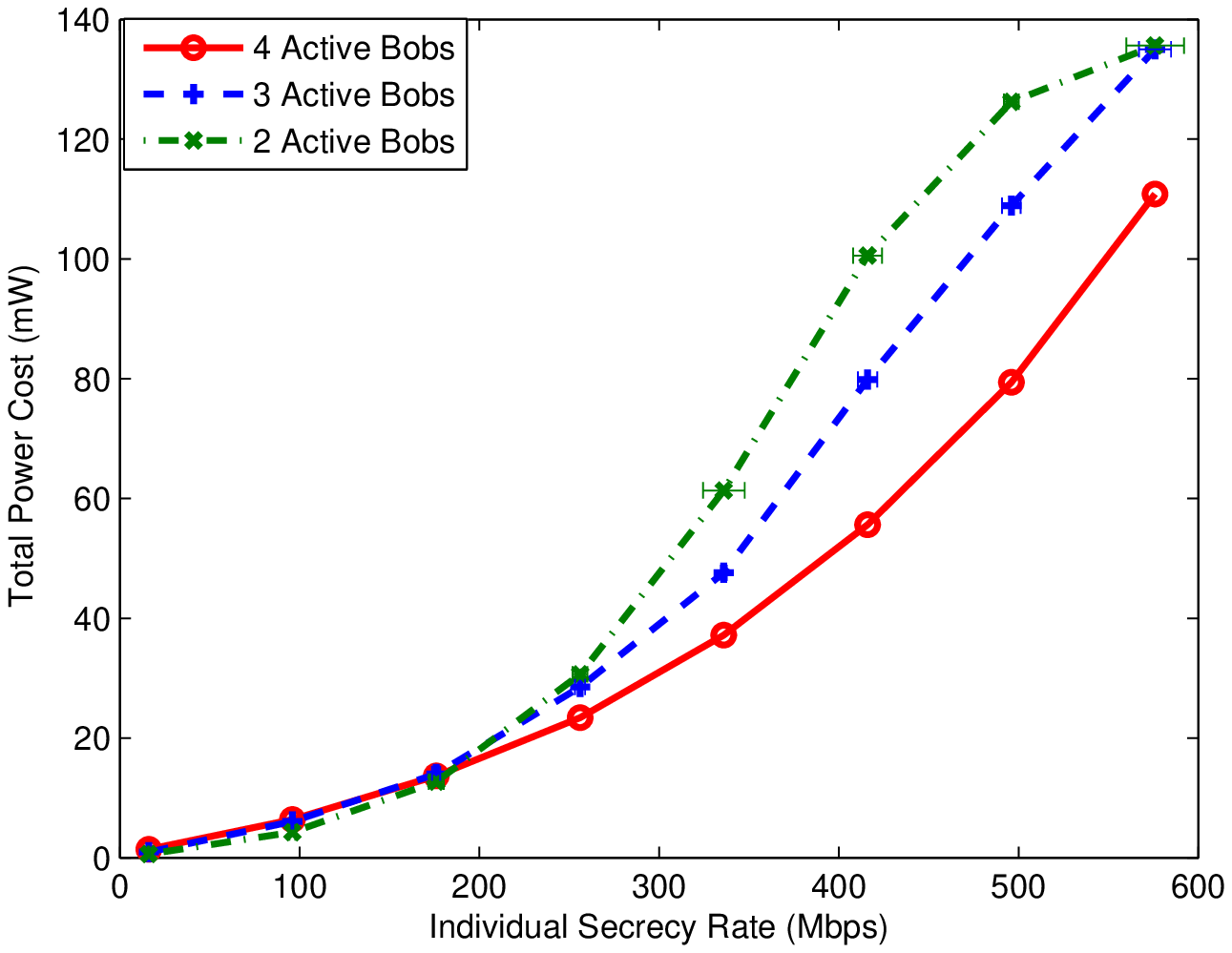}
        \caption{Independently located Bobs with nearby Eves ($\rho = 0.9$)}
        \label{fig:4b}
    \end{subfigure}    
    \caption{Total power consumption vs. individual secrecy rate with known ECSI where $K = 4$ and $L = 3$.}
    \label{fig:4}
\end{figure}

We compare the results of such a scheduling scheme in Figs. \ref{fig:4a} and \ref{fig:4b}.
Different Bobs are selected for each communication block, as a different topology is created each time.
Fig. \ref{fig:4a} is obtained for the case of known ECSI, where $ \rho = 0$, $\alpha = 0$, $K = 4$, and $L = 3$.
Number of scheduled (active) Bobs, $T$, at a given time is showed in the legend.
Note that the x-axis represents the average individual secrecy rate over 4 Bobs. (Horizontal bars indicate standard deviation of the achieved rates. We note that the number of repetitions, which is 3000, is enough to have almost equal rates in the long term.)
In the low and high power regime, the proposed scheduling scheme (transmitting to fewer Bobs) outperforms the regular one (transmitting to all of Bobs).
However, between some threshold points, transmitting to $3$ Bobs consumes less power than transmitting to $2$ or $4$ Bobs.
On the other hand, Fig. \ref{fig:4b} shows that the regular scheme always has a better performance, when Eves are located around Bobs, and the channel correlation coefficient between Bobs and the corresponding Eves is equal to $0.9$.
In this case, selecting the closest Bobs to Alice makes the performance worse, as Eves are also close to Alice due to the assumption that they are in the vicinity of Bobs.
Similar results are obtained for the unknown ECSI case.

\section{Conclusion}

In this paper, we considered the scenario where a transmitter sends $K$ independent confidential data streams, intended to $K$ legitimate receivers in the presence of $L$ eavesdroppers. With the knowledge that the security applications require guard zones around receivers up to 19 wavelengths, we proposed using RxFJ along with TxFJ.
That way, even if an eavesdropper has a highly correlated channel with that of any legitimate receiver and is able to cancel out TxFJ, RxFJ keeps facilitating confidentiality for the information signals. To be able to send RxFJ from the receivers, we considered FD receivers.
These receivers are capable of partial/complete self-interference suppression. We used zero-forcing beamforming technique not only to remove the TxFJ interference at intended receivers but also to hide the information signals from the unintended receivers. We showed how to design practical precoders for information signals and TxFJ signals.
We formulated a minimum power allocation problem to the information, TxFJ, and RxFJ signals under certain secrecy rate requirements.
We solved this problem with/without the knowledge of eavesdropper's CSI. The results showed that using RxFJ together with TxFJ increases the system performance in multiuser MISO systems, especially when the eavesdropper channels are correlated with that of the legitimate receivers.  

Throughout this paper, only Bobs had FD capabilities. We note that if Eves had such FD capabilities as well, they would be able to send jamming signals to decrease the SINR at Bobs, while simultaneously eavesdropping the information messages over the same frequency. Problems that arise from this model are left for future studies. We also initiated a study of scheduling schemes (here, based on the distance between Alice and Bobs) to further decrease the power consumption. The results showed that under certain conditions, different scheduling methods increase the performance. The effect of other scheduling strategies in the context of secret communications will be reported elsewhere.
 

\bibliography{PaperDatabase}
\bibliographystyle{IEEEtran}

\end{document}